\documentclass[numreferences]{kluwer}

\begin{document}                                      
\begin{article}
\begin{opening}         
  \title{The Cauchy Relations in Linear
    Elasticity Theory
    } \author{Friedrich W.\ \surname{Hehl}} \runningauthor{F.W.\ Hehl
    and Y.\ Itin} \runningtitle{On the Cauchy Relations}
  \institute{Institute for Theoretical Physics, University of Cologne,
    50923 K\"oln, Germany {\tt hehl@thp.uni-koeln.de}} \author{Yakov
    \surname{Itin}}\institute{Institute of Mathematics, Hebrew
    University of Jerusalem,  Jerusalem 91904, Israel;\\
    Jerusalem College of Engineering, Jerusalem
    91035, Israel \\{\tt itin@math.huji.ac.il}}

\date{11 June 2002, {\it file elasticity/cauchy12.tex}}

\begin{abstract}
  In linear elasticity, we decompose the elasticity tensor into two
  irreducible pieces with 15 and 6 independent components,
  respectively. The {\it vanishing} of the piece with 6 independent
  components corresponds to the Cauchy relations. Thus, for the first
  time, we recognize the group-theoretical underpinning of the Cauchy
  relations.
\end{abstract}

\end{opening}           

\section{Linear elasticity}

In linear elasticity theory for homogeneous bodies, the stress tensor
$\sigma^{ij}=\sigma^{ji}$, with $i,j,\dots=1,2,3$, is related to the
strain tensor\footnote{We use the notation of Schouten
  \cite{Schouten}. Symmetrization over two indices is denoted by
  parentheses, $A_{(ij)}:=(A_{ij} +A_{ji})/2!$, antisymmetrization by
  brackets, $B_{[ij]}:=(B_{ij} -B_{ji})/2!$. The analogous is valid
  for more indices, as, e.g., in (\ref{rest**}): $c^{i[jk]l}=
  (c^{ijkl} -c^{ikjl})/2!$.  If one or more indices are exempt from
  (anti)sym\-metrization, they are enclosed by vertical bars, as,
  e.g., in (\ref{rest'*}), $c^{(i|kl|j)}=(c^{iklj}+c^{jkli})/2! $, or,
  more complicated, in (\ref{firstx}), $c^{(i|(kl)|j)}=
  (c^{i(kl)j}+c^{j(kl)i})/2=(c^{iklj} +c^{ilkj}+ c^{jkli}+c^{jlki})/4
  $. Symmetrization over more than 2 indices is, by definition, the
  normalized sum over all possible permutations of the indices
  involved.  Thus, $c^{(ijkl)}:=(c^{ijkl}+c^{jikl} +c^{ijlk}+ \mbox{21
    more terms})/4!$, see (\ref{first}). The totally antisymmetric
  Levi-Civita symbol is denoted by $\epsilon_{ijk}$ and
  $\epsilon^{lmn}$, with $\epsilon_{123}=\epsilon^{123}=+1$; for all
  even (odd) permutations of $1,2,3$, we have $+1$ ($-1$), otherwise
  $0$. The partial derivative $\partial/\partial x^k$ we denote by
  $\partial_k$.}
$\varepsilon_{kl}=\varepsilon_{lk}=\partial_{(k}\,u_{l)}$ by Hooke's
law,
\begin{equation}\label{hooke}
\sigma^{ij}=c^{ijkl}\,\varepsilon_{kl}\,,
\end{equation} 
see, for instance,  \shortcite{Marsden} or \shortcite{Sommerfeld}.  
Here $u_l$ is the displacement
field and $c^{ijkl}$ the constant 4th rank elasticity tensor.

Since stress and strain are symmetric tensors, the elasticity tensor
obeys the symmetries
\begin{equation}\label{symmetries}
c^{ijkl}=c^{jikl}=c^{ijlk}\,.
\end{equation}
A symmetric 2nd rank tensor has 6 independent components. Therefore,
by collecting the indices $i$ and $j$ into an index pair and $k$ and
$l$ likewise, the elasticity tensor can be thought of as a $6\times 6$
matrix with 36 independent components.

Usually one assumes that the stress tensor can be derived from an elastic
potential $W$ (also called ``strain energy function''), i.e.,
\begin{equation}\label{pot}
  \sigma^{ij}=\frac{\partial W}{\partial\varepsilon_{ij}}\,,\qquad{\rm
    or}\qquad c^{ijkl}=\frac{\partial^2\,W}{\partial
    \varepsilon_{ij}\,\partial \varepsilon_{kl}}\,.
\end{equation}
Then the first and the last pair of indices of $c^{ijkl}$ commute,
\begin{equation}\label{paircom}
  c^{ijkl}= c^{klij}\,, 
\end{equation}
and the $6\times 6$ matrix is symmetric and carries only $36-15=21$
independent components. Thus, as is well known, the elasticity tensor
has $21$ independent components.

\section{Generally covariant decomposition of the elasticity 
  tensor}

It is remarkable that the elasticity tensor can be decomposed under
the group of general coordinate transformations (diffeomorphisms) into
two irreducible pieces. Locally this amounts to the application of the
3-dimensional linear real group $GL(3,R)$. Via an analysis of Young
tableaux, one can determine that a totally symmetric tensor of rank
$r$ in $n$ dimensions has, expressed as binomial coefficient,
${n+r-1\choose r}$ independent components, see 
\shortcite{Schouten}.  Accordingly,
\begin{equation}\label{first}
  ^{(1)}c^{ijkl}:=c^{(ijkl)}
\end{equation}
has ${3+4-1\choose 4}=15$ independent components. Because of the
symmetries (\ref{symmetries}) and (\ref{paircom}), we find
\begin{equation}\label{firstx}
 c^{(i|(kl)|j)}=c^{i(kl)j}=c^{(i|kl|j)}\,,
\end{equation}
and we can go over to the more condensed formula
\begin{equation}\label{firsty}
^{(1)}c^{ijkl}=\frac 13(c^{ijkl}+c^{iklj}+ c^{iljk})\,.
\end{equation}
Alternatively, we can write
\begin{eqnarray}\label{first'}
  ^{(1)}c^{ijkl}&=& 
 \frac 13\left(c^{ijkl}+2c^{(i|(kl)|j)} \right)\\ &=&\frac
  13\left(c^{ijkl}+2c^{i(kl)j} \right)=\frac
  13\left(c^{ijkl}+2c^{(i|kl|j)} \right)\,.
\end{eqnarray}

The surviving irreducible piece of $c^{ijkl}$, with $21-15=6$
independent components, encodes the {\it excess} of the elasticity tensor
over its totally symmetric piece,
\begin{equation}\label{rest}
 ^{(2)}c^{ijkl}:=c^{ijkl}-\,^{(1)}c^{ijkl}\,,
\end{equation}or, by means of (\ref{first'}),
\begin{eqnarray}\label{rest'}
  ^{(2)}c^{ijkl}&=& \frac13\left(2c^{ijkl}-c^{iklj}-c^{ilkj}\right) =
  \frac23\left(c^{ijkl}-c^{(i|(kl)|j)}\right)
  \\&=&\frac23\left(c^{ijkl}-c^{i(kl)j}\right)=
  \frac23\left(c^{ijkl}-c^{(i|kl|j)}\right)\,.\label{rest'*}
\end{eqnarray}
We can also express the right hand side in terms of antisymmetric
terms,
\begin{equation}\label{rest**}
^{(2)}c^{ijkl}=\frac 23\left(c^{i[jk]l}+c^{i[jl]k}\right)=
\frac 23\left(2c^{i[jk]l}+c^{i[kl]j}\right)\,.
\end{equation}
While the tensor $c^{i[jl]k}$ has already been introduced in elasticity
theory \shortcite{Fosdick}, this tensor itself is not an irreducible
piece of the elasticity tensor. In particular, it does not inherit the
symmetries of $ c^{ijkl}$.

The irreducible pieces $^{(\alpha )}c^{ijkl}$ have, as a rule, the same
algebraic symmetries as $c^{ijkl}$. We list them here for later use:
\begin{equation}\label{symmetries(2)}
  ^{(\alpha)}c^{[ij]kl}=\,^{(\alpha)}c^{ij[kl]}=0\,,\qquad
  ^{(\alpha)}c^{ijkl}=\, ^{(\alpha)}c^{klij}\,,\qquad\alpha=1,2\,.
\end{equation}
Moreover, because of (\ref{rest}) and (\ref{first}), we have
\begin{equation}\label{cyclic}
  ^{(2)}c^{(ijkl)}=0\,.
\end{equation}
 Therefore finally we have the decomposition
\begin{equation}\label{dec}
\mathbf{  c^{ijkl}=\, ^{(1)}c^{ijkl}+\, ^{(2)}c^{ijkl}}
\end{equation}or
\begin{equation}\label{dec1}
21= 15\oplus 6\,.
\end{equation}
An analysis by means of Young tableaux guarantees that this
decomposition is unique and cannot be continued successively. Similar
group theoretical methods were applied for the decomposition of the
electromagnetic constitutive tensor of the vacuum
$\chi^{\mu\nu\kappa\lambda}$, a 4th rank tensor in 4 dimensions, see
\shortcite{Hehl}.

\section{Mapping $\mathbf{ ^{(2)}c^{ijkl}}$ 
  to a symmetric 2nd rank tensor}

Since $ ^{(2)}c^{ijkl}$ has 6 independent components, it is to be
expected that it can be mapped to a symmetric 2nd rank tensor.
Clearly, the Levi-Civita symbol $\epsilon_{ijk}$ has to be used in
this context. However, the Levi-Civita symbol can only be applied to
antisymmetric index pairs, otherwise information may get lost.
Consequently, we should derive, out of $ ^{(2)}c^{ijkl}$, another
tensor which has two pairs of antisymmetric indices. The obvious
method is to antisymmetrize suitably:
\begin{equation}\label{tildedef}
  ^{(2)}\hat{c}^{ijkl}:=\,^{(2)}{c}^{[i|[jk]|l]}=
  \,^{(2)}{c}^{i[jk]l}={c}^{i[jk]l} \,.
\end{equation}
For the derivation of the last equality, we employed (\ref{rest}).
Because of (\ref{symmetries(2)})$_1$, the antisymmetrization in the
definition (\ref{tildedef}) cannot be applied to the first or the last
pair of indices. Thus, modulo different conventions, this is a unique
procedure. Note also that the index pairs in (\ref{tildedef}) commute.
Thus, we have the symmetries
\begin{equation}\label{symmetries(3)}
^{(2)}\hat{c}^{ijkl}=-^{(2)}\hat{c}^{ljki}=-^{(2)}\hat{c}^{ikjl}=\,
^{(2)}\hat{c}^{lkji}=\,
^{(2)}\hat{c}^{klij}\,.
\end{equation}
In contrast to (\ref{symmetries(2)})$_1$,
\begin{equation}\label{symmetries(4)}
  ^{(2)}\hat{c}^{[ij]kl}\neq 0\,,\qquad \,^{(2)}
  \hat{c}^{ij[kl]}\neq 0\,.
\end{equation}

In order to show that $ ^{(2)}\hat{c}^{ijkl}$ is equivalent to $
^{(2)}c^{ijkl}$ and carries, in particular, the same number of
independent components, namely 6, we have to resolve (\ref{tildedef})
with respect to $ ^{(2)}c^{ijkl}$. We substitute
${c}^{i[jk]l}=\,^{(2)}\hat{c}^{ijkl}$,
$c^{i[jl]k}=\,^{(2)}\hat{c}^{ijlk}$, and
${c}^{i[kl]j}=\,^{(2)}\hat{c}^{iklj}$ into (\ref{rest**}) and obtain
\begin{equation}\label{tildedef3}
  ^{(2)}c^{ijkl}= \frac 43\, ^{(2)}\hat{c}^{ij(kl)}\,=\frac 23\left(2\,
    ^{(2)}\hat{c}^{ijkl}+\, ^{(2)}\hat{c}^{iklj} \right)\,.
\end{equation}
This is the inverse of (\ref{tildedef}) and proves the equivalence of $
^{(2)}c^{ijkl}$ and $ ^{(2)}\hat{c}^{ijkl}$.

We remember that $^{(2)}\hat{c}^{ijkl}$ is antisymmetric in $il$
and in $jk$, see (\ref{symmetries(3)}). Therefore the dual of the
``excess'' tensor $^{(2)}\hat{c}^{ijkl}$ turns out to be
\begin{equation}\label{ansatz}
\Delta_{mn}:=\frac
14\epsilon_{mil}\epsilon_{njk} \,^{(2)}\hat{c}^{ijkl}\,.
\end{equation}
The tensor $\Delta_{mn}$ is symmetric and has 6 independent
components. Indeed, by using (\ref{symmetries(3)}), we find
\begin{eqnarray}\label{sym}
  \Delta_{nm}&=&\frac 14\epsilon_{nil}\epsilon_{mjk}
  \,^{(2)}\hat{c}^{ijkl} =\frac 14\epsilon_{mil}\epsilon_{njk}
  \,^{(2)}\hat{c}^{jilk}= -\frac 14\epsilon_{mil} \epsilon_{njk}
  \,^{(2)}\hat{c}^{jlik}\nonumber \\&=& \frac
  14\epsilon_{mil}\epsilon_{njk} \,^{(2)}\hat{c}^{klij}=
  \Delta_{mn}\,.
\end{eqnarray}

With the help of (\ref{tildedef}), Eq.(\ref{ansatz}) can be also
expressed in terms of the original elasticity tensor:
\begin{eqnarray}\label{ansatz*'}
  \mathbf{ \Delta_{mn}}&=&\mathbf{\frac 14\epsilon_{mil}\epsilon_{njk}
    \,{c}^{ijkl}}\\ & =&\frac 14\epsilon_{mil}\epsilon_{njk}\,
  ^{(2)}c^{ijkl} \,.\label{ansatz*'*}
\end{eqnarray}
The last equation is valid, since $^{(1)}c^{ijkl}$ is a totally
symmetric tensor and thus
\begin{equation}\label{ansatz*''}
  \frac 14\epsilon_{mil}\epsilon_{njk} \,^{(1)}{c}^{ijkl}=0\,.
\end{equation}
In other words, the operator $\frac 14\epsilon_{mil}\epsilon_{njk}$
extracts the dual of the 2nd irreducible piece $^{(2)} c^{ijkl}$ from
$c^{ijkl}$, it is a type of a projection operator.  Another projection
operator is, of course, the symmetrizer $(....)$ in (\ref{first}).

Formula (\ref{ansatz*'}) has been given earlier in the literature by
Hauss\"uhl \shortcite{Hauss}. However, our derivation is new.

For the inversion of (\ref{ansatz*'}), it is very convenient to use
the generalized Kronecker symbols, see  \shortcite{Sokol}.
We multiply the left hand side of (\ref{ansatz*'}) by two 3-dimensional
epsilons:
\begin{eqnarray}\label{inverse}
  \epsilon^{pqm}\epsilon^{r\!sn}\,
  \Delta_{mn}&=&\frac{1}{4}\epsilon^{pqm}\epsilon^{r\!sn}
  \epsilon_{mil}\epsilon_{njk}\,c^{ijkl} =
  \frac{1}{4}\,\delta^{pq}_{il}\delta^{r\!s}_{jk}\,c^{ijkl}\nonumber\\ 
  &=& \frac{1}{2}\,\delta^{pq}_{il}\,c^{i[r\!s]l}= c^{[p|[r\!s]|q]}=
  \,^{(2)}\hat{c}^{pr\!sq} \,,\quad\mbox{q.e.d..}
\end{eqnarray}

\section{Cauchy relations}

Around 1830, in the early days of modern elasticity theory, Navier,
Poisson, Cauchy, and others set up molecular models for elastic
bodies, see Todhunter \cite{Todhunter}. If they prescribed, inter
alia, certain properties to the interaction forces between these
molecules that build up the macroscopic body, then they found 6
constraining relations between the 21 components of the elasticity
tensor. Nowadays, they are called Cauchy relations.  A theory obeying
the Cauchy relations and thus carrying only 15 independent elastic
constants was called {\it rari-constant} theory, the general case with
21 elastic constants {\it multi-constant} theory. For decades, there
was a scientific battle between both camps. For the special case of
isotropy, merely 1 independent elastic constant survives the Cauchy
relations, see Sec.5.

More recent discussions of the Cauchy relations can be found, e.g., in
\shortcite{Brillouin}, \shortcite{Hauss*}, \shortcite{Hauss}, or
\shortcite{Love}.  A lattice-theoretical approach to the elastic
constants shows, see \shortcite{Leibfried}, that the Cauchy relations
are valid provided (i) the interaction forces between the atoms or
molecules of a crystal are central forces, as, e.g., in rock salt,
(ii) each atom or molecule is a center of symmetry, and (iii) the
interaction forces between the building blocks of a crystal can be
well approximated by a harmonic potential. In most elastic bodies this
is not fulfilled at all, see \shortcite{Hauss*}. However, a study of
the {\it violations} of the Cauchy relations yields important
information about the intermolecular forces of elastic bodies.

If we take the fundamental meaning of the Cauchy relations for
granted, then it is near at hand to assume that the vanishing of
$^{(2)}c^{ijkl}$ corresponds to the 6 Cauchy relations:
\begin{equation}\label{cauchy1}
^{(2)} c^{ijkl}=0\qquad\quad\mbox{(Cauchy relations)\,.}
\end{equation}
According to the Young tableau technique, there is no other way to
split off 6 components from the elasticity tensor in a generally
covariant and irreducible way. Incidentally, Eq.(\ref{cauchy1}) is,
due to (\ref{ansatz*'*}), equivalent to $\Delta_{ij}=0$.

Let us show that (\ref{cauchy1}) really encompasses the Cauchy
relations. According to (\ref{cauchy1}), (\ref{rest}), and
(\ref{first}), the elasticity tensor is now totally symmetric:
$c^{ijkl}=c^{(ijkl)}$. Consequently,
\begin{equation}\label{cauchy1*}
\mathbf{c^{ijkl}=c^{ikjl}}\,.
\end{equation}
In this form, the Cauchy relations can be found in \shortcite{Hauss}.
In ``longhand'', the nontrivial components of Eq.(\ref{cauchy1*}) read
\begin{eqnarray}\label{longhand}
&&  c^{1122}=c^{1212}\,,\quad c^{1133}=c^{3131}\,,\quad
  c^{2233}=c^{2323}\,,\\&& c^{1123}=c^{1231}\,,\quad
  c^{2231}=c^{2312}\,,\quad c^{3312}=c^{3123}\,.
\end{eqnarray}
In the $6\times 6$ notation, $ 1\cong 11\,, \,2\cong 22\,,\,3\cong
33\,,\, 4\cong 23\,,\, 5\cong 31\,,\, 6\cong 12$, we find the
alternative formulae
\begin{eqnarray}\label{longhand'}
 && c^{12}=c^{66}\,,\quad c^{13}=c^{55}\,,\quad
  c^{23}=c^{44}\,,\\ && c^{14}=c^{56}\,,\quad
  c^{25}=c^{46}\,,\quad c^{36}=c^{45}\,.
\end{eqnarray}
These coefficients coincide with those of Love in \shortcite{Love}
p.100, Eq.(13). But certainly, formula (\ref{cauchy1*}) is much more
compact and easier to remember.

\section{Isotropy}

Let us test our results for the rather trivial case of isotropy. For
an isotropic body, the elasticity tensor can be expressed in terms of
the metric tensor $g^{ij}$ as
\begin{equation}\label{iso}
  c^{ijkl}=\lambda\,g^{ij}g^{kl}+\mu\left(g^{ik}g^{lj}+g^{il}g^{jk}
  \right)\,,
\end{equation}
with the Lam\'e moduli $\lambda$ and $\mu$, see 
\shortcite{Marsden}.  Consequently,
\begin{equation}\label{first''}
  ^{(1)} c^{ijkl}= (\lambda+2\mu)\,g^{(ij}g^{kl)}=
  \frac{\lambda+2\mu}{3}\left(g^{ij}g^{kl}+
    g^{ik}g^{lj}+g^{il}g^{jk}\right)\,.
\end{equation}
The 2nd irreducible piece can be conveniently determined by
substituting (\ref{first''}) directly into (\ref{rest}):
\begin{equation}\label{second'}
  ^{(2)} c^{ijkl}= \frac{\lambda-\mu}{3}\left(2g^{ij}g^{kl}-
     g^{ik}g^{lj}-g^{il}g^{jk}\right)\,.
\end{equation}
Furthermore,
\begin{equation}\label{inver}
  ^{(2)}\hat{c}^{ijkl}={c}^{i[jk]l}=(\lambda-\mu)\,g^{i[j}g^{k]l}\,.
\end{equation}
The corresponding 2nd rank tensor (\ref{ansatz*'*}) turns out to be
\begin{equation}\label{piso}
  \Delta_{mn}=\frac{\lambda-\mu}{4}\,\epsilon_{mil}\epsilon_{njk}\,
  g^{i[j}g^{k]l}=\frac{\lambda-\mu}{4}\,\epsilon_{m}{}^{jl}\epsilon_{njl}\,.
 \end{equation}
Accordingly,
\begin{equation}\label{piso*8}
 \Delta_{ij}= \frac{\lambda-\mu}{2}\,g_{ij}\,.
\end{equation}
Thus, for an isotropic body, the Cauchy relations translate into
$\lambda=\mu$. Then the elasticity tensor depends only on one modulus.
For most isotropic elastic bodies this is certainly not fulfilled.

\appendix

An analogous decomposition can be made for a 4th rank tensor
$d^{ijkl}$ which lacks the ``Onsager symmetry'' (\ref{paircom}) but
which still carries the symmetries (\ref{symmetries})
\begin{equation}\label{symmetries'}
  d^{ijkl}=d^{jikl}=d^{ijlk}\,.
\end{equation}
The {\it piezooptical tensor} is of such a type, see 
\shortcite{Hauss} p.269, for instance. The first irreducible piece is
defined as in (\ref{first}):
\begin{eqnarray}\label{totalsym}
  ^{(1)}d^{ijkl}:=d^{(ijkl)}&=&\frac
  16\left(d^{ijkl}+d^{ikjl}+d^{iljk} \right.\nonumber\\ 
  &&\;\, \left. +d^{klij}+d^{jlik}+d^{jkil}\right)\,.
\end{eqnarray}
Then we quantify the violation of the pair symmetry by
\begin{equation}\label{nono}
  ^{(3)}d^{ijkl}:= \frac{1}{2}\left(d^{ijkl}- d^{klij}\right)\,
\end{equation} and the rest as 
\begin{eqnarray}\label{rest*}
  ^{(2)}d^{ijkl}&:=&d^{ijkl}-\,^{(1)}d^{ijkl}-\, ^{(3)}d^{ijkl}\\ 
  \nonumber &=&\frac 13\left[d^{ijkl}+d^{klij}-\frac 12
    \left(d^{ikjl}+d^{jlik} \right)- \frac 12\left(d^{iljk}+d^{jkil}
    \right)\right]\,.
\end{eqnarray}
Finally we have the decomposition
\begin{equation}\label{dec*}
  d^{ijkl}=\, ^{(1)}d^{ijkl}+\, ^{(2)}d^{ijkl}+\, ^{(3)}d^{ijkl}
\end{equation}or
\begin{equation}\label{dec1*}
36= 15\oplus 6 \oplus 15\,,
\end{equation}
which is irreducible.

In analogy to (\ref{ansatz}), one can map $^{(2)}d^{ijkl}$ to a 2nd
rank tensor:
\begin{equation}\label{ansatz*}
  \Delta_{mn}:=\frac 14\epsilon_{mil}\epsilon_{njk} \,^{(2)} d^{ijkl}\,.
\end{equation}
It is again symmetric, that is, $\Delta_{[mn]}=0$, as can be shown in
an analogous way as in (\ref{sym}). Moreover, since
$^{(1)}d^{ijkl}=\,^{(1)}d^{(ijkl)}$, we have $\frac
14\epsilon_{mil}\epsilon_{njk} \,^{(1)} d^{ijkl}=0$. As for the third
part,
\begin{eqnarray}\label{third}
  \frac 14\epsilon_{mil}\epsilon_{njk} \,^{(3)}d^{ijkl}&=& \frac
  14\epsilon_{mli}\epsilon_{nkj} \,^{(3)}d^{lkji}= \frac
  14\epsilon_{mil}\epsilon_{njk} \,^{(3)}d^{klij}\nonumber\\ &=&-\frac
  14\epsilon_{mil}\epsilon_{njk} \,^{(3)}d^{ijkl}=0\,.
\end{eqnarray}
Consequently,
\begin{equation}\label{ansatz**}
  \Delta_{mn}=\frac 14\epsilon_{mil}\epsilon_{njk} \, d^{ijkl}\,.
\end{equation}

Imposing the additional symmetry $ d^{ijkl}= d^{klij}$, that is,
$^{(3)} d^{ijkl}=0$, leads back to the results found for the
elasticity tensor $c^{ijkl}$.

\acknowledgements We would like to thank E.\ Jan Post (Los Angeles)
for a hint to look into the Cauchy relations: ``...It is possible that
Cauchy did something similar what you guys are doing...'' in
electrodynamics. We are grateful to Siegfried Hauss\"uhl (Cologne) for
interesting discussions and to Roger Fosdick (Minneapolis) for his
communications.  Both authors appreciate the help and the support by
Shmuel Kaniel (Jerusalem).

\end{article}
\end{document}